\documentclass[num-refs]{wiley-article}
\papertype{Research Article}
\usepackage{hyperref}
\usepackage{booktabs}
\usepackage[table]{xcolor}  
\usepackage[utf8]{inputenc}
\usepackage{url}
\usepackage{microtype}
\usepackage{subfig,subfloat}
\usepackage{tikz}
\usepackage{tikzscale}
\usetikzlibrary{external}
\usepackage{dcolumn}
\usepackage{algorithm}
\usepackage{algpseudocode}
\usepackage[shortcuts]{extdash}
\newcolumntype{g}{D{.}{.}{-1}}
\newcommand{\mc}[1]{\multicolumn{1}{c}{#1}}
\usepackage{pgfplots}
\usepgfplotslibrary{external}
\usetikzlibrary{external}

\definecolor{bblue}{HTML}{4F81BD}
\definecolor{rred}{HTML}{C0504D}
\definecolor{ggreen}{HTML}{9BBB59}
\definecolor{ppurple}{HTML}{9F4C7C}

\newcommand*{\codepoint}[1]{\texttt{U+\MakeUppercase{#1}}}
\newcommand*{\codepointrange}[2]{\texttt{U+\MakeUppercase{#1}\ldots\MakeUppercase{#2}}}
\newcommand*{\hexrange}[2]{\texttt{0x\MakeUppercase{#1}\ldots\MakeUppercase{#2}}}
\newcommand*{\utfbinary}[2]{\texttt{\begingroup\color{darkgray}#1$\vert$\endgroup#2}}

\newcommand*{\restartrowcolors}{%
  \ifhmode\unskip\fi
  \vadjust{%
    \global\rownum=0 %
  }%
}

\usepackage[binary-units,per-mode=symbol]{siunitx}

\title{Transcoding Billions of Unicode Characters per Second with SIMD Instructions}

\runningauthor{Daniel Lemire and Wojciech Muła}

\author[1\authfn{1}]{Daniel Lemire}
\affil[1]{DOT-Lab Research Center, Universit\'e du Qu\'ebec (TELUQ)}
\author[2\authfn{2}]{Wojciech Muła}
\affil[2]{0x80.pl, Wrocław, Poland}

\corraddress{Daniel Lemire, DOT-Lab Research Center, Universit\'e du Qu\'ebec (TELUQ), Montreal, Quebec, H2S 3L5, Canada}
\corremail{daniel.lemire@teluq.ca}
\fundinginfo{Natural Sciences and Engineering Research Council of Canada, Grant Number: RGPIN-2017-03910}

\begin{document}

\maketitle

\begin{abstract}In software, text is  often represented using Unicode formats (UTF-8 and UTF-16). 
We frequently have to convert text from one format to the other, a process called transcoding. Popular transcoding functions are slower than state-of-the-art disks and networks.
These transcoding functions make little use of the single-instruction-multiple-data (SIMD) instructions available on commodity processors. 
By designing transcoding algorithms for SIMD instructions,  we multiply the speed of transcoding  on current systems (x64 and ARM). To ensure reproducibility, we make our software freely available as an open source library.

\end{abstract}


\section{Introduction}

Unicode assigns a  number from 0~to \num{1114112}---i.e., a \textit{code point}---to  every character in every language. These code points are stored as bytes in a computer using one of several character encoding formats (e.g., UTF-8, UTF-16, UTF-32). 
UTF-8 is the most popular encoding format~\cite{rfc3629}: it is the default for HTML, XML and JSON documents. UTF-8 uses between 1~and 4~bytes to encode each character. It is also an extension of ASCII: ASCII bytes are automatically valid UTF-8 bytes.
UTF-16 is another popular format~\cite{rfc2781}, used by default in Java 
and .NET\@. UTF-16 uses either 2~or 4~bytes per character. 
For internal processing within software functions, there is also the UTF-32 encoding format which uses 4~bytes per character. 
All of these formats require validation: an arbitrary sequence of bytes may not be valid~\cite{keiser2020validating}.


Transcoding and validating encoded text is relatively efficient (e.g., \SIrange{0.5}{1}{\gibi\byte\per\second}), but a single processor cannot match the maximal bandwidth between node instances in cloud computing (e.g., \SI{3.3}{\gibi\byte\per\second}) or the sequential throughput of fast disks (e.g., \SI{5}{\gibi\byte\per\second})~\cite{bar2008}. Slower functions tend to use more power than needed~\cite{mitra2013use}. Given how common transcoding may be, even a small efficiency gain could be worth the effort. Furthermore,  it is generally useful to know how fast an important operation could be, as an engineering constraint. 

Most commodity processors have single-instruction-multiple-data (SIMD) instructions working on wide registers (e.g., 128-bit). These instructions can accelerate software and reduce power usage when they achieve the same tasks with fewer instructions. We often refer to algorithms and functions using SIMD instructions as \emph{vectorized}.\footnote{Vector processing does not require SIMD instructions in general but in does in our context~\cite{eichenberger2004vectorization}.} SIMD instructions  have been applied to the character transcoding problem. Yet we found relatively few references in the scientific literature to vectorized transcoding algorithms (\S~\ref{sec:relatedwork}). 

Our main contribution is a set of techniques that transcode any input from either UTF-8 or UTF-16 to the other encoding, with validation if needed. We implemented our techniques as part of a freely available C++ open source library. Our library works on common operating systems (Linux, macOS, Windows) and CPU architectures (x86, ARM, POWER). It compiles to a binary library using only tens of kilobytes. Our approaches often multiply the performance of existing non-SIMD solutions while being generally faster and more practical than existing SIMD-based solutions. 

\section{Related Work}
\label{sec:relatedwork}

Cameron~\cite{cameron2008case} proposed a patented  UTF-8 to UTF-16 transcoder using SIMD instruction on \emph{bit streams}. To construct a bit stream, one  may transform byte-oriented character stream data into eight parallel bit streams~\cite{cameron2007u8u16}. From 128~bytes of inputs, eight 128-bit registers are constructed corresponding to the least significant bit of each of the 128~bytes, the second least significant bit and so forth.
Cameron describes the cost of these transformations as acceptable. On an Intel Core 2, they report that the forward transform (from UTF-8 bytes to bitstream)  requires 1.6~cycles/byte whereas the inverse transform (from bitstream to UTF-16) requires approximately 4.0~cycles/byte. Thus the general overhead due to the bitstream transforms is over 5~cycles per byte, at least on some processors and in some cases. Once the data is in bit stream, an efficient data parallel algorithm  converts the 8~bit streams from UTF-8 into 16~bit streams representing UTF-16 characters. In an intermediate phase, the UTF-16 bit stream contains unused bits which must be removed or compacted.

Inoue et al.~\cite{Inoue2008} proposed a patented UTF-8 to UTF-16 SIMD-accelerated transcoder (for PowerPC). It accelerated processing by up to a factor of three compared to a non-accelerated baseline. It does not validate the input and it is limited to 3-byte inputs.
It consists in a single loop. At each iteration of the loop, the next eight~UTF-8 characters are identified. Because 4-byte characters are excluded, the eight characters use between 8~and 24~bytes. They use the length in bytes of each of the eight~characters (1, 2~or 3) to construct an index---an integer value no longer than $3^8$. They then load two 16-byte registers from the next 32~bytes of the input source.  They use the computed index to permute the bytes into two new 16-byte registers: one contains either the ASCII values, or the least significant 6-bit of a character, while the other register contains the rest of the bit values (between 0~and 10~bits). To permute the bytes, they require a lookup table thus occupying $3^8\times 32$~bytes or about \SI{105}{\kibi\byte}. 
The ability to permute bytes within a register in an arbitrary manner, based on a lookup table, is a key ingredient of their approach as well as of most other similar techniques, including our own.
E.g., to reverse the order of bytes, we could load up the register with the indexes in reverse order ($15,14,\ldots,3,2,1,0$) and apply it to a 16-byte register.
In practice, most SIMD instruction sets have a corresponding fast instructions (e.g., \texttt{pshufb} under x64 processors) and it has a broad range of applications~\cite{lemire2018stream,mula2018faster,mula2018fasterpop,zhao2015general,lemire2017upscaledb,stepanov2011simd}.
Inoue et al.\ also suggest using fast code paths for common sequences like ASCII strings. We present a simplified version of their algorithm in Algorithm~\ref{algo:inoue}. 
We refer the interested reader to the original article for details.

\begin{algorithm}[htb]
\begin{algorithmic}[1]
\Require  an array of bytes $b$, containing values in $[0,256)$
\Require  an array of 16-byte values  $c$ (in $[0,2^{16})$ with sufficient capacity
\Require two precomputed 6561-element tables containing 16-byte values ($\mathrm{pattern1}$ and $\mathrm{pattern2}$ ).
\State $p\leftarrow 0$, $q\leftarrow 0$
\While {$p + 32 < \mathrm{length}(b)$}
\If{$b[p], b[p+1],\ldots  b[p+7]$ are ASCII}
\State $c[q] = b[p], c[q+1] = b[p+1], \ldots, c[q+7] = b[p+7]$
\State $p\leftarrow p + 8$, $q\leftarrow q + 8$
\Else 
\State $v \leftarrow $ load 32~bytes from $b[p]$ to $b[p+31]$ 
\State $g \leftarrow 0$
\For{$i = 0, \ldots, 7$}
\State lookup index $b[p+i] \div 2^5$ in table $[1, 1, 1, 1, 1, 1, 2, 3]$ to get length $l\in [1,3]$

\State $g \leftarrow 3 * g + (l-1)$
\State $p \leftarrow p + l$
\EndFor
\State $p_1 \leftarrow \mathrm{vector\_load\_16bytes}(\mathrm{pattern1}[g])$
\State $p_2 \leftarrow \mathrm{vector\_load\_16bytes}(\mathrm{pattern2}[g])$
\State  $v_1 \leftarrow \mathrm{vector\_permute}(v, v_1)$, $v_2 \leftarrow \mathrm{vector\_permute}(v, v_2)$
\State $v'_1 \leftarrow \mathrm{vector\_shift\_left}(v_1,4)$, $v''_1 \leftarrow \mathrm{vector\_shift\_left}(v_1,6)$
\State $v_1 \leftarrow \mathrm{vector\_select}(v'_1,v''_1,\mathrm{0x0FFF})$
\State $v_2 \leftarrow \mathrm{vector\_and}(v_2,\mathrm{0x7F})$
\State $v \leftarrow \mathrm{vector\_or}(v_1,v_2)$
\State Treating $v$ as eight 16-bit values, write them to $c[q], \ldots, c[q+7]$ and increment $q$ by 8.
\EndIf
\EndWhile
 \State \textbf{Return}: The counter value $q$ indicating the number of 16-bit words written.
\end{algorithmic}
\caption{%
\label{algo:inoue}
Inoue et al.~\cite{Inoue2008} UTF-8 to UTF-16 transcoder. }
\end{algorithm}

We find it interesting that the only formally published SIMD-accelerated UTF-8 to UTF-16 transcoders that we could find~\cite{cameron2008case,Inoue2008} were published the same year (2008). It followed the availability of commodity processors with powerful SIMD instructions (e.g.,  the IBM POWER6 in 2007, the Intel Core microarchitecture in 2006).

We are not aware of any further work in the scientific literature. We could not find work on UTF-16 to UTF-8 transcoding.
 More recently, Keiser and Lemire~\cite{keiser2020validating} proposed SIMD-based UTF-8 validation directly on byte streams. They report 10~times the performance of competitive alternatives. We are unaware of any formal proposal to validate UTF-16 text using SIMD instructions but it is a relatively simpler problem that may not warrant an extensive study.
 
\begin{table}\centering
\begin{tabular}{lcccl}\toprule
                                      & UTF-8 to UTF-16 & UTF-16 to UTF-8   & validation & platforms  \\\midrule
Cameron~\cite{cameron2008case} (2008) & yes & no & yes & x64 and POWER \\
Inoue et al.~\cite{Inoue2008} (2008) & partial & no  & no & POWER \\
Goffart~\cite{Goffart2012} (2012) &yes & no & yes & x64 (SSE4) \\  
Gatilov~\cite{stgatilov} (2019) &yes & yes  & yes (optional) & x64  \\  \midrule
our work &yes & yes  & yes (optional) & x64, 64-bit ARM \\  
     \bottomrule
\end{tabular}
\caption{\label{table:summary} Related work with respect to SIMD-based Unicode transcoding (excluding work limited to ASCII fast paths).}
\end{table}

Many transcoding functions use SIMD instructions, but they often do so only to accelerate specific code paths such as ASCII sequences. We are focused on the more general problem of transcoding potentially challenging cases (e.g., Asiatic text).
We found two other relevant engineering efforts that apply SIMD instructions more broadly with recent x64 (Intel and AMD) processors:

\begin{itemize}
    \item In 2012, Goffart~\cite{Goffart2012} proposed a fast decoder capable of transcoding UTF-8 to UTF-16. They reported competitive results compared with the UTF-8 to UTF-16 transcoder from Cameron~\cite{cameron2008case}. Unlike Inoue et al.~\cite{Inoue2008}, they can detect errors. Their work is specific to the x64 SSE4 instruction set. Characters outside the basic multilingual plane---requiring more than 3~bytes per character---are handled in a non-SIMD manner.
    They load 16~bytes at each step.
    They have a fast path for sequences of 16~ASCII characters. Otherwise, 
    from these 16~bytes, they identify the start of each UTF-8 character. Instead of using a table-based approach (like Inoue et al.~\cite{Inoue2008} and others), they use several \emph{byte shifts} and \emph{blend} instructions to move the byte data.
    They consume 14, 15, or 16~bytes per iteration and always  produce two 16-byte output registers.
    Their routine contains relatively many instructions but few branches and no lookup table.
    \item More recently (2019), Gatilov~\cite{stgatilov} offered a software library that provides both UTF8 to UTF-16 and UTF-16 to UTF-8 transcoding, with or without validation.
    A downside of Gatilov's approach is the size of the lookup tables: about \SI{2}{\mebi\byte} for the UTF-8 to UTF-16 transcoder and about \SI{16}{\kibi\byte} for the UTF-16 to UTF-8 transcoder. The \SI{2}{\mebi\byte} table might be a concern for some engineers. Unlike some other related work, Gatilov covers transcoding in the two directions (from UTF-8 to UTF-16 and back). They provide extensive documentation as well as a rich set of parameters, documentation and tests. Gatilov's code only accelerates the basic multilingual plane so that strings involving 4-byte UTF-8 characters may incur performance penalties.
\end{itemize}
There is undoubtedly other related work found online, but both Goffart and Gatilov provided a complete implementation in elegant C++, with documentation. We expect that they represent the state-of-the-art.
We provide a summary of the related work in Table~\ref{table:summary}.

\section{The UTF-8 and UTF-16 Formats}

There are slightly more than 1~million Unicode characters. We can think of them as integer  values---called code points---from 0~to \num{1114112}.\footnote{A Unicode glyph may be represented by several Unicode characters.} We could represent them all using 21-bit machine words but current processors have machine words spanning 8~bits, 16~bits, 32~bits or 64~bits. The UTF-32 format relies on 32-bit words to represent characters, but it is a wasteful format from a storage point of view. In practice, software increasingly represents Unicode characters using either UTF-8 or UTF-16. They are both variable-length formats---the length in bytes of characters can differ. As their names suggest, the UTF-8 format relies on 8-bit words whereas the UTF-16 format relies on 16-bit words.

We represent character values using the hexadecimal notation. The first Unicode character is \codepoint{0000} and the last one is \codepoint{10FFFF}. The 2048~characters in the range \codepointrange{D800}{DFFF} are omitted and forbidden.
Both the UTF-8 and UTF-16 formats concisely represent frequenly used characters (see Table~\ref{table:utfsummary}). ASCII characters require one byte with UTF-8 and two bytes with UTF-16. UTF-16 can represent all characters---except for the supplemental characters such as emojis---using two bytes. The UTF-8 format uses two bytes for Latin, Hebrew and Arabic alphabets. Asiatic characters (including Chinese and Japanese) require three UTF-8 bytes. Both UTF-8 and UTF-16 require 4~bytes for the supplemental characters. These formats are normalized: there is only one way to encode a given Unicode character.\footnote{Though there are different ways to encode the same glyph.}
Not all sequences of bytes can be interpreted as valid UTF-8 or UTF-16. We must validate sequences of bytes, especially if they come from an external or untrusted source.

\begin{table}\centering
\begin{tabular}{lcc}\toprule
character range & UTF-8 bytes & UTF-16 bytes  \\\midrule
 ASCII (\codepointrange{0000}{007F})    & 1  &  2\\
 latin (\codepointrange{0080}{07FF})    & 2  & 2  \\
 asiatic (\codepointrange{0800}{D7FF} and \codepointrange{E000}{FFFF})    & 3  & 2  \\
 supplemental \codepointrange{010000}{10FFFF}  & 4  & 4  \\\bottomrule
\end{tabular}
\caption{\label{table:utfsummary} Number of bytes used by the UTF-8 and UTF-16 formats for range of Unicode characters}
\end{table}

UTF-8 encodes values in sequences of bytes (between 1~and 4~bytes) starting with a leading byte. The most significant bits of the leading byte determine the length in bytes of the sequence. If the most significant bit of the leading byte is zero, we have an ASCII character: the sequence has a length of one byte. If the three most significant bits are \texttt{110}, then we have a two-byte sequence; if the most significant bits are \texttt{1110} then we have a three-byte sequence and finally, if the most significant bits are \texttt{11110} then we have a 4-byte sequence. For two-byte, three-byte and four-byte characters, all bytes following the leading byte are continuation bytes: their most significant bits must be \texttt{10}. Not counting the required most significant bits, we can encode the character value using 7~bits, 11~bits, 16~bits and 21~bits depending on the length of the sequence in bytes. In UTF-8, the encoding is always big-endian meaning that the most significant bits are first encoded in the leading byte, then in the first continuation byte and so forth. Thus, for example, the character \codepoint{93E1} (or \texttt{0b1001001111100001} in binary) is encoded as a three-byte sequence \utfbinary{1110}{1001}, \utfbinary{10}{001111},\utfbinary{10}{100001}. To validate UTF-8 bytes, we must enforce all of the following rules:
\begin{enumerate}
\item The five most significant bits of any byte cannot be all ones. E.g., the byte values \texttt{0b11111000} can never be part of a UTF-8 stream. Any large byte value such as \texttt{0b11111111} is similarly forbidden.
\item The leading byte must be followed by the right number of continuation bytes (0, 1, 2~or 3). 
\item Conversely, a continuation byte must be preceded by a leading byte (up to 3~bytes prior).
\item The decoded character must be larger than \codepoint{7F} for two-byte sequences, larger than \codepoint{7FF} for three-byte sequences, and larger than \codepoint{FFFF} for four-byte sequences. 
\item The decoded character must be less  than \codepoint{110000} (\num{1114112} in decimal). 
\item The  character must not be in the range \codepointrange{D800}{DFFF}. 
\end{enumerate}
These rules are exhaustive.

The UTF-16 format is maybe simpler. Characters with values in 
\codepointrange{0000}{D7FF} and \codepointrange{E000}{FFFF} are stored as 16-bit values, as is. There are two common ways to store 16-bit values in bytes: little endian, with the least significant bits in the first byte,  and big endian, with the most significant bits in the first byte. For this reason, UTF-16 comes in two flavors: little endian and big endian encoding. To differentiate between the two formats, it is possible to start the character stream with a byte-order mask. The two bytes \texttt{0xff 0xfe} indicate a little-endian format whereas the two bytes \texttt{0xfe 0xff} indicate a big-endian format. It is recommended to omit the byte-order mask when the format has been labeled~\cite{rfc2781} or when it is otherwise known.
The endianness of the data often matches the 
endianness of the underlying system. Popular systems such as Windows are little-endian systems and there are relatively few big-endian systems left. Thus the little-endian format is often assumed by default. Whereas most characters require a single  16-bit word in UTF-16, 
the characters in the range
\codepointrange{010000}{10FFFF} require two 16-bit words, called a surrogate pair. The first word must be in  the range \hexrange{d800}{dbff} while the second word must be in the range \hexrange{dc00}{dfff}. We then construct the character value by first setting its 10~least significant bits to the 10~least significant bits of the second 16-bit word, and then setting the next 10~least significant bits to the 10~least significant bits of the first 16-bit word. We get a value in the range \hexrange{00000}{fffff} (in $[0,2^{20})$). Finally, we add \texttt{0x10000} to get a value in \hexrange{10000}{10ffff} as required.
As long as a stream of characters does not include characters in the range \codepointrange{010000}{10FFFF}, UTF-16 validation consists merely in checking that no 16-bit value in the range 
\hexrange{d800}{dfff} appears. Otherwise, we must check that a words in the range \hexrange{d800}{dbff} is always followed by a word in the range \hexrange{dc00}{dfff}, and conversely. In practice, surrogate pairs are relatively uncommon. Thus validating UTF-16 may merely involve checking for the absence of 16-bit words in the range \hexrange{d800}{dfff}.

\begin{figure}
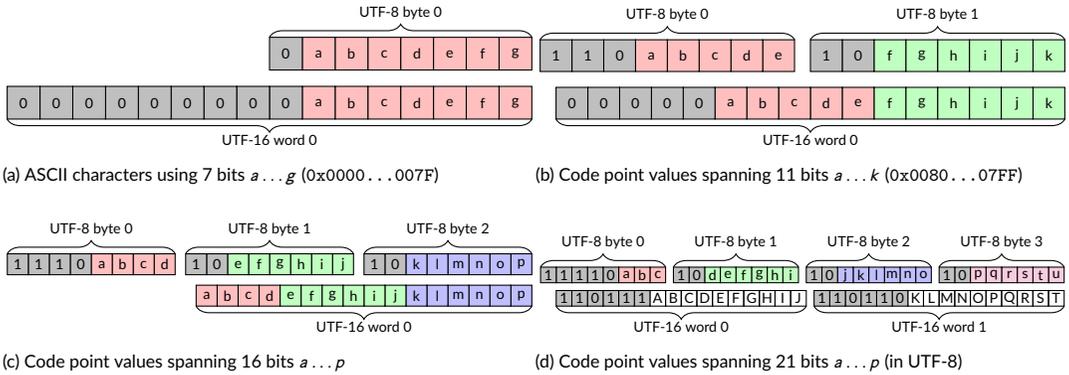
\centering\scriptsize
\subfloat[ASCII characters using 7~bits $a\ldots g$ (\hexrange{0000}{007f})\label{fig:subascii}]{
\includegraphics[width=0.49\textwidth]{images/simdutf-mapping-1byte.tikz}
}
\subfloat[Code point values spanning 11~bits $a\ldots k$ (\hexrange{0080}{07ff})\label{fig:sub2}]{
\includegraphics[width=0.49\textwidth]{images/simdutf-mapping-2bytes.tikz}
}\\
\subfloat[Code point values spanning  16~bits $a\ldots p$\label{fig:sub3} ]{
\includegraphics[width=0.49\textwidth]{images/simdutf-mapping-3bytes.tikz}
}
\subfloat[Code point values spanning 21~bits $a\ldots p$ (in UTF-8)\label{fig:sub4}]{
\includegraphics[width=0.49\textwidth]{images/simdutf-mapping-4bytes.tikz}
}
\caption{\label{fig:correspondence}Bit-by-bit correspondence between UTF-8 and UTF-16 encodings}
\end{figure}

Fig.~\ref{fig:correspondence} illustrates the work necessary to transcode characters from one format to the other. For ASCII characters (\hexrange{0000}{007f}), we  just need to add or remove a zero byte (Fig.~\ref{fig:subascii}). For the other characters in the basic multilingual plane---using 2~bytes or 3~bytes in UTF-8 format---we just need to move bits (e.g., by shift operations), see Figs.~\ref{fig:sub2} and~\ref{fig:sub3}. For characters outside of the basic multilingual plane---requiring 4~bytes in either UTF-8 or UTF-16---more work is needed (see Fig.~\ref{fig:sub4}).

\section{Transcoding UTF-8 to UTF-16}

We can implement transcoding from UTF-8 to UTF-16 in various ways. A finite-state machine, possibly with additional fast paths, is competitive~\cite{Steagall2018,Hoehrmann}. We may also apply a brute-force branching approach: we look at each incoming byte, check that is it is a leading byte, and branch on the expected number of continuation bytes (from 0~to 3).

One difficulty when transcoding variable-length format is that 
we must realign them. Characters require a
different number of bytes in UTF-8 than they do in UTF-16. Thus when converting characters
from one format to another, we have to move bits
at variable locations in both the input and output buffers.
We can do so easily when working character-by-character, but with a vectorized approach that processes many characters at once, we need to be able to move the byte values all at once within a block.

As described in \S~\ref{sec:relatedwork}, practically all commodity software with SIMD instructions have fast instructions to shuffle or permute bytes within a SIMD register arbitrarily according to a sequence of indexes. When storing these series of indexes precomputed in tables, we can quickly align the character sequences. 
Instead of precomputing the sequences of indexes (sometimes called shuffle masks), we could also 
compute them as we go, but we expect that it comes at a significant performance penalty.
Thus, to vectorize the processing beyond just the detection of some easy cases, we may use a table-based approach. In such an approach, we code into a table the parameters---including the shuffle masks---necessary to process a variety of sequences of incoming bytes.

Our  accelerated  algorithm (see Algorithm~\ref{algo:coreutf8toutf16component}) processes data by reading up to 12~input UTF-8 bytes at a time and by writing between 6~and 12~bytes in UTF-16. It can be implemented efficiently with tables. From the input bytes, we can quickly determine the start of each new character: any byte that is not continuation byte must be a leading byte.
It follows that we can also find the end of each character: the byte preceding the start of a new character must be the end of the previous character.
We can compute a bitset (i.e., a 12-bit word) indicating the location of the end of each character.
 A vectorized byte-by-byte comparison followed by a reduction suffices. We can use this 12-bit bitset as a key in a 1024-entry table. Each entry in the table gives us the number of bytes that will be consumed (following Algorithm~\ref{algo:coreutf8toutf16component}) as well as an index into another table where we find shuffle masks. We sort the indexes: the first 64~index values (in $[0,64)$) indicate the first case in Algorithm~\ref{algo:coreutf8toutf16component}: 6~characters spanning between one and two bytes.
 Index values in $[64,145)$ represent the second case. Index values in $[145,209)$ represent the third and final case. This main table requires 1024~entries each spanning two bytes---one byte indicating the number of consumed bytes and another byte for the index---for a total of  \SI{2}{\kibi\byte}. 
 The shuffle mask can then be applied to the 12~input bytes to form a vector register than can be transformed efficiently.
We need 209~shuffle masks, each spanning 16~bytes for a total of \SI{3.3}{\kibi\byte}. In exchange for a small performance cost, we could compress these tables. In practice, only a fraction of these tables is accessed when processing a given input string. For example, when processing a French text, we may only rarely encounter characters requiring more than 2~bytes.

Working in units of 12~bytes is somewhat arbitrary. The same approach would work with 6~bytes or 32~bytes. Small blocks require more iterations and may be less efficient, especially since all modern SIMD architectures have registers capable of storing 16~bytes. Wider blocks seem more efficient but  require larger tables. Tables that fail to fit in the fastest CPU cache (L1)  lead to high-latency loads and poor performance. Large tables increase either the memory usage or the binary size of an application. The best block size should depend on system's architecture but we expect that 12~bytes is a sensible platform-agnostic choice.

\begin{algorithm}
\begin{algorithmic}[1]
\Require  an array of 12~bytes $b$ containing characters in UTF-8 format
\If{$b$ starts with 6~characters spanning one or two bytes each}
\State convert the corresponding bytes (between 6~and 12~bytes) to a  12-byte UTF-16 sequence containing six characters (see Fig.~\ref{fig:utf8case1})
\ElsIf{$b$ starts with 4~characters spanning one, two or three bytes each}
\State convert the corresponding bytes  (between 6~and 12~bytes) to an 8-byte UTF-16 sequence containing four characters  (see Fig.~\ref{fig:utf8case2})
\Else
\State Convert the bytes of two characters (between 5~and 8~bytes) into  UTF-16 bytes, producing between four and eight~bytes  (see Fig.~\ref{fig:utf8case3}) 
\EndIf
\end{algorithmic}
\caption{%
Core table-based component of our UTF-8 to UTF-16 transcoders. \label{algo:coreutf8toutf16component}}
\end{algorithm}

Processing a UTF-8 stream using a 12-byte approach might be inefficient in some cases. Thus we use our core routine as part of the more general Algorithm~\ref{algo:nonvalidatingutf8toutf16transcoder}. We load input bytes in blocks of 64~bytes. We can efficiently detect whether they are all ASCII bytes, in which case we apply a fast path. Otherwise we compute once a bitset indicating the end of the characters, and apply Algorithm~\ref{algo:coreutf8toutf16component} in a loop (at least five times).
A strength of our approach is that it is relatively content-agnostic: we handle 1-byte, 2-byte, 3-byte and 4-byte UTF-8 characters as part of the same code path. 

To validate the input bytes, we apply the Keiser-Lemire approach which already works in chunks of 64~bytes~\cite{keiser2020validating}. The Keiser-Lemire approach is efficient: it requires few instructions. Furthermore, we only need to validate the UTF-8 input when it is not ASCII\@. Thus, maybe surprisingly, we can get validation at a relatively modest runtime cost.

\begin{figure}\centering
\includegraphics[width=0.99\textwidth]{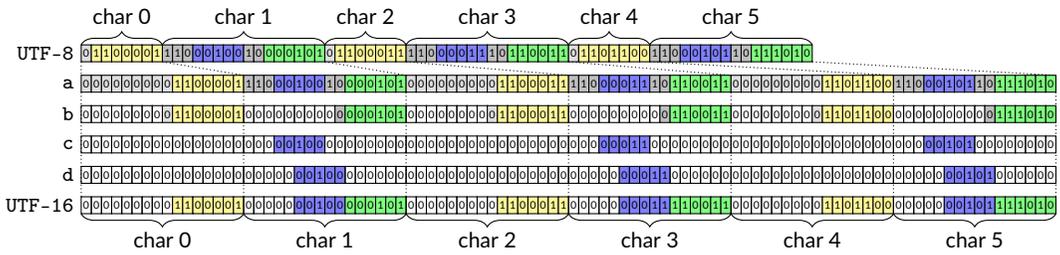}
\caption{\label{fig:utf8case1}Vectorized UTF-8 to UTF-16 
transcoding of six  ASCII and 2-byte UTF-8 input characters.
Based on the character sizes, we load an appropriate shuffle mask and place the characters 
in separate 16-bit words of a vector register (\texttt{a}). We isolate the seven least significant 
bits in each byte (vector \texttt{b}). For 2-byte characters, the $7^{\mathrm{th}}$~bit 
is always zero. We isolate the data bits from the leading UTF-8 byte (vector \texttt{c}) 
and we shift it right by 2~bits (vector \texttt{d}). 
Finally, we merge the bits from vectors \texttt{a} and \texttt{d} which yields the final 
six UTF-16 characters that can be directly written out.
}
\end{figure}

\begin{figure}\centering
\includegraphics[width=0.99\textwidth]{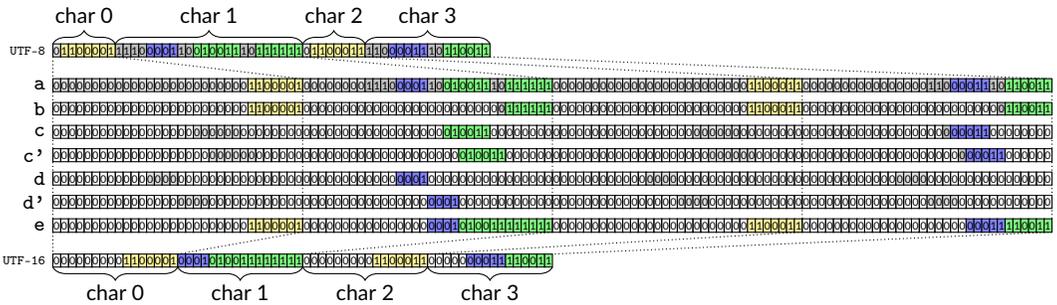}
\caption{\label{fig:utf8case2}
Vectorized UTF-8 to UTF-16 
transcoding of four input characters from the basic multilingual plane. The input sequence contains ASCII, 2-byte and 3-byte UTF-8 characters. Based on the character sizes, we load an appropriate shuffle mask and place characters in separate 32-bit words of vector \texttt{a}. We isolate ASCII words (vector \texttt{b}). We isolate data bits from the leading byte (for 2-byte UTF-8 chars) or the second byte (for 3-byte UTF-8 characters). This step produces the vector \texttt{c}, which is then shifted right by two bits (vector \texttt{c'}). Likewise, we isolate data bits from the leading byte of 3-byte UTF-8 characters (vector \texttt{d}) and shift it right by four bits (vector \texttt{d}). Finally, we merge all intermediate vectors \texttt{b}, \texttt{c'} and \texttt{d'} into the vector \texttt{e}. At this point the lower 16~bits of the 32-bit words contain UTF-16 character, thus the higher 16~bits are discarded in a final step.}
\end{figure}

\begin{figure}\centering
\includegraphics[width=0.80\textwidth]{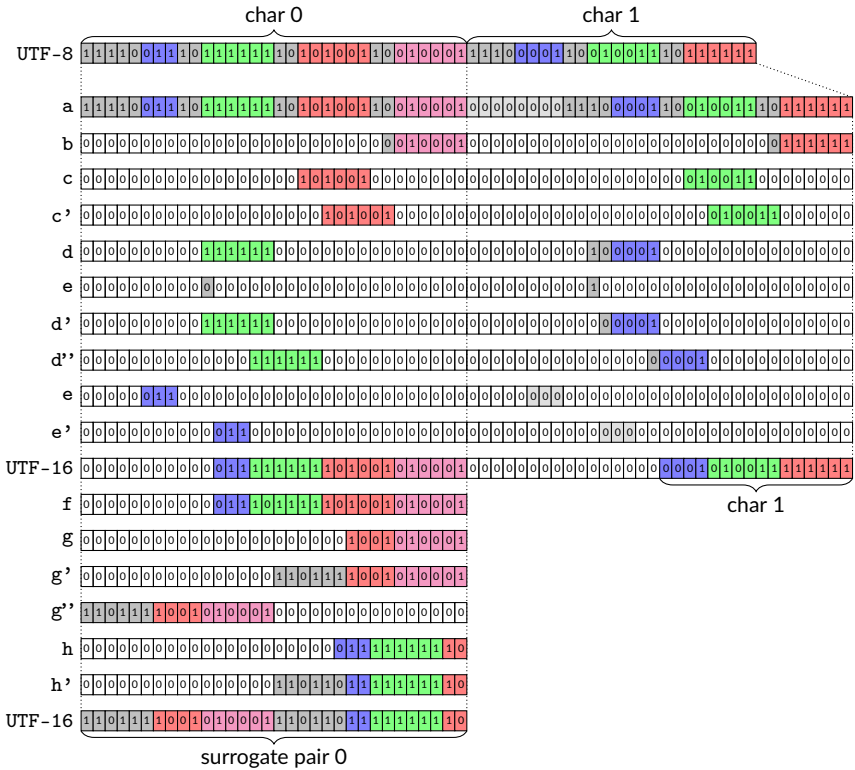}
\caption{\label{fig:utf8case3}
Vectorized UTF-8 to UTF-16 
transcoding of two input characters. The input sequence may contain any UTF-8 character: ASCII, 2-byte, 3-byte and 4-byte characters. Based on the character sizes we load an appropriate shuffle mask and place characters in separate 32-bit words within vector \texttt{a}. We isolate the 7~least significant bits (vector \texttt{b}). We isolate the data bits from the second byte into vector \texttt{c} and shift it right by 2~bits (vector \texttt{c'}). Likewise, we isolate data bits from the fourth byte in vector \texttt{e} and shift it right by 6~bits (vector \texttt{e'}). More work is required to properly isolate data bits from the third byte. In the case of 4-byte characters, the third byte has 6~data bits, but in case of 3-byte characters, this byte is the leading byte and contains only four bits. In the latter case, the isolated field has two fixed bits \texttt{10}. We use the $7^{\mathrm{th}}$~bit from third byte to conditionally reset the extra 1~bit~at the $6^{\mathrm{th}}$~bit~position. We isolate that bit and shift it right by one (vector \texttt{e}). For 4-byte characters that $7^{\mathrm{th}}$~bit is always 0, for 3-byte character, it is always always 1. Next we perform an exclusive-or operation on vectors \texttt{d} and \texttt{d} which yields vector \texttt{d'}. Finally that last vector is shifted right by 4~bits (vector \texttt{d''}). After merging vectors \texttt{b}, \texttt{c'} and \texttt{d''}, we obtain a UTF-16 characters in least significant 16~bits for ASCII, 2- and 3-byte UTF-8 characters. We store this result and continue handling 4-byte characters which yield surrogate pairs. We follow the algorithm given in the UTF-16 speficiation. First, from the input 21-bit word, we subtract \texttt{0x10000} (vector \texttt{g}). Then we isolate 10~least significant bits (\texttt{h}), we complete with the six bits \texttt{110111} (\texttt{h'}) and shift left by 16~bit (\texttt{h''}). Then we isolate the 10~most significant bits by shifting 32-bit words by 10~bits right (\texttt{i}) and compete with the six bits \texttt{110110} (\texttt{i'}). The bitwise or of vectors \texttt{h''} and \texttt{i'} produces the final surrogate pairs.
}
\end{figure}

\begin{algorithm}[htb]
\begin{algorithmic}[1]
\Require  an array of bytes $b$, containing values in  UTF-8 format
\Require  an array of 16-byte values  $c$ (in $[0,2^{16})$ with sufficient capacity (e.g., $2*\mathrm{length}(b)$)
\State $p\leftarrow 0$, $q\leftarrow 0$
\While {$p + 64 < \mathrm{length}(b)$}
\State Load 64~bytes in vector $v$
\State Compare the signed 8-bit byte values in $v$ with $-65$:  all bytes less than $-65$ using two complement's notation are continuation bytes.
\If{there is no continuation byte}
\State Treating $v$ as 64~8-bit values, write them to $c[q], \ldots, c[q+64]$ and increment $p$  and $q$ by 64.
\Else 
\State Store the 64~Boolean values `is not a continuation byte' in a 64-bit word $z$ so that the least significant bit matches the first byte encountered.
\State Shift the word $z$ by one bit left (i.e., remove the least significant bit).
\State $m\leftarrow p + 64 - 12$
\While {$p<m$}
\State Apply Algorithm~\ref{algo:coreutf8toutf16component} to process six, four or two characters based on the least significant 12~bits of $z$, consuming $k$~input bytes and writing the converted characters to $c$, incrementing $q$ as needed 
\State increment $p$ my $k$, shift the word $z$ by $k$~bits left
\EndWhile
\EndIf
\EndWhile
 \State \textbf{Return}: The counter value $p$ indicating how many input bytes were processed and the counter value $q$ indicating the number of 16-bit words written.
\end{algorithmic}
\caption{%
Vectorized non-validating UTF-8 to UTF-16 transcoder. \label{algo:nonvalidatingutf8toutf16transcoder}}
\end{algorithm}

The algorithm described so far is efficient but we can do slightly better in some cases.
Before calling Algorithm~\ref{algo:coreutf8toutf16component} from  Algorithm~\ref{algo:nonvalidatingutf8toutf16transcoder}, we can check with a few branches whether the next few bytes match common patterns. We check the following three patterns by examining the bitmask (noted $z$ in Algorithm~\ref{algo:nonvalidatingutf8toutf16transcoder}).
First, we check if the next 16~bytes are ASCII bytes---we recognize such a case when all 16~bits are set. Second, we check whether the next 16~bytes are two-byte characters---we recognize such a case from the 16-bit bitset value \texttt{0xaaaa}. Finally, we check whether the next 12~bytes are made of three-byte characters. For each case, we have an efficient routine. Inoue et al.\ alluded to such fast paths~\cite{inoue2008fast}.

Our approach reads the data in blocks of 64~bytes (Algorithm~\ref{algo:nonvalidatingutf8toutf16transcoder}). We fall back on a conventional approach to process the remaining  bytes (1~to 63~bytes).

\section{Transcoding UTF-16 to UTF-8}

When converting from UTF-16 into UTF-8, we have to handle the following scenarios:

\begin{enumerate}
    \item words in the range \codepointrange{0000}{007f} yield a single ASCII byte,
    \item words in the range \codepointrange{0080}{07ff} yield two UTF-8 bytes,
    \item words in ranges \codepointrange{0800}{d7ff} and \codepointrange{e000}{ffff} yield three  UTF-8 bytes.
\item Surrogate pairs, made of two 16-bit words in the range  \codepointrange{d800}{dfff}, yield four UTF-8 bytes.
\end{enumerate}

Our algorithm reads a block of input bytes in a SIMD register. The number of bytes read depends on the size of the SIMD registers of the platform (e.g., 16~bytes on ARM NEON and 32~bytes on x64 AVX2). To simplify the exposition, we assume that we have 16-byte registers and that we load up to eight UTF-16~characters. When we have a wider register (32~bytes), we effectively apply the same algorithm, but treat the wider register as a block of two 16-byte registers.

Iterating through the input, register-by-register, we branch on the content of the register:
\begin{enumerate}
\item If all 16-bit words in the loaded SIMD register are in the range \codepointrange{0000}{007f}, we use a fast routine to convert the 16~input bytes into eight equivalent ASCII bytes.
\item If all 16-bit words are in the range \codepointrange{0000}{07ff}, then we use a specialized routine to produce sequences of one-byte or two-byte UTF-8 characters. In this case, we can always convert the 16~input bytes into up to 16~output bytes. 
We use a table-based approach: given an 8-bit bitset which indicates which 16-bit words are ASCII, we load a byte value from a table indicating how many bytes will be written, and a 16-byte shuffle mask. Thus we need a table spanning $256\times 17=4352$~bytes.
\item If all 16-bit words are in the ranges \codepointrange{0000}{d7ff}, \codepointrange{e000}{ffff}, we use another specialized  routine to produce sequences of one-byte, two-byte and three-byte UTF-8 characters. In this scenario, we may need to produce up to 24~bytes from the incoming 16~bytes, so that the output may not fit in a single SIMD register.
We also use a table-based approach.
We treat the input as a pair of 8-byte registers. For each, 
we produce an 8-bit bitset which indicates which 16-bit words are ASCII, and which would produce 2-byte UTF-8 words. We then a shuffle mask and a byte-value to indicate how many bytes are written.
Thus we need another 4352-byte~table.
\item Otherwise, when we detect that the input register contains at least one part of a surrogate pair, we fall back to a conventional code path. Although it is possible to vectorize the conversion algorithm---and our initial implementation had such a vectorized approach, we found it may require large lookup tables or other undesirable compromises.
We only need to validate the input in the presence of such a potential surrogate pair. Furthermore, it is only in this particular scenario that we may not be able to consume the entire 16-byte input---when it ends with the beginning of a surrogate pair.
\end{enumerate}
Overall, we require two tables spanning a total of 8704~bytes.
See Algorithm~\ref{algo:utf16toutf8component}.

\begin{algorithm}
\begin{algorithmic}[1]
\Require load 16~bytes
\If{all bytes are in the range \codepointrange{0000}{007f}}
\State convert the corresponding 8~ASCII characters to eight UTF-8 bytes 
\ElsIf{all bytes are in the range  \codepointrange{0000}{07ff}}
\State convert the eight characters to between 8~and~16~UTF-8 bytes
\ElsIf{all bytes are in the ranges  \codepointrange{0000}{d7ff}, \codepointrange{e000}{ffff}}
\State convert the eight characters to between 8~and~24~UTF-8 bytes
\Else
\State Convert between four and eight characters to UTF-8 using a conventional path, consuming between 14~and~16~bytes.
\EndIf
\end{algorithmic}
\caption{%
Our UTF-16 to UTF-8 transcoders. \label{algo:utf16toutf8component}}
\end{algorithm}

Our two routines, for the range \codepointrange{0000}{07ff} and for the ranges \codepointrange{0000}{d7ff}, \codepointrange{e000}{ffff}, are similar. In both cases, the number of output bytes varies. However, in the first scenario, the output  fits in a SIMD register. For the second scenario, the output may be larger than a single SIMD register. Thus we need to cast the input 16-bit words into 32-bit words prior to the actual conversion to UTF-8. 
Conversion in both cases is performed in a similar manner. We split the bits of the input words into potential UTF-8 bytes: we place 3, 4, 5~bits in the leading byte and 6~bits in continuation bytes. We identify how many output bytes are produced by each input 16-byte  word. We then complete the bit layout in each byte to follow the UTF-8 standard---we add leading the bits \texttt{11110}, \texttt{1110}, \texttt{110} or \texttt{10} to the appropriate bytes. Finally, we \emph{compress}  the words into a continuous sequence of bytes using the shuffle mask loaded from the table.

As in the UTF-8 to UTF-16 routine, we may end the function with few unprocessed input bytes. We terminate the processing with a conventional routine. Because it runs over few bytes, it does not contribute much to the total running time.

\section{Experiments}

We seek to validate that we can provide high  transcoding speed  from UTF-8 to UTF-16 and back on current commodity processors using SIMD instructions. To establish such a claim, we need to compare against competitive software libraries on realistic data.

\subsection{Software}

All of our code, including the benchmarking software, our own implementations, and competitive implementations are freely available.\footnote{\url{https://github.com/simdutf/simdutf}} All code is built with the \texttt{cmake}  tool in release mode.
The data input is not aligned in a particular manner. We include unit tests to ensure that our code is correct. We avoid buffer overruns: our tests run without error with memory sanitizers and under tools such as \texttt{valgrind}.

Our own code is written using a high-level C++ approach which allows us to easily support multiple processor instruction sets (e.g., ARM NEON, AVX2, SSE2). We rely on the compiler's optimizer. We also did not try to tune our code to a specific compiler nor do we use assembly.

We benchmark how quickly we can convert  data files from UTF-8 to UTF-16 and back, while working entirely in memory, in a single thread. We  repeat the task \num{2000}~times (on the x64 platform) and \num{10000}~times (on the ARM platform), measuring the time for each conversion. The distribution of timings is approximately log-normal. We use the minimum timing after comparing it with the average. We verify automatically that the difference between the minimum and the average is small (less than 1\%). In all tests, we attempt to minimize system calls and  memory allocations. There is no disk access during benchmarking. 

We report performance results in characters per second as opposed to other more conventional measures such as bytes or megabytes per second. The benefit of measuring by the number of characters is that it is format oblivious: a given string has the same number of characters whether it is in UTF-8 or UTF-16 format. Thus it makes it possible to compare the relative performance of transcoding the same string from UTF-8 to UTF-16 and from UTF-16 to UTF-8. 

While UTF-16 has two subformats (little endian and big endian), we focus on the little endian format for simplicity. Supporting the big-endian UTF-16 format given a little-endian transcoder requires little effort, especially with SIMD instructions. The ARM NEON instruction set has specialized instructions (e.g., \texttt{rev16}) but it is always possible to use byte shuffling instructions (e.g.,  \texttt{pshufb} under x64 systems).

We compare our work with the following SIMD-accelerated competitors:
\begin{itemize}
\item To our knowledge, Inoue et al.~\cite{Inoue2008} did not publish their software and their work was focused on the POWER platform. Therefore, for comparison purposes, we reimplemented their approach for both x64 (using SSE) and ARM (using NEON) platforms. As per the description in their manuscript, we also include a fast path when ASCII sequences are encountered. We expect our implementation is representative of the core Inoue et al.\ algorithm though we are certain that their implementation was more finely tuned.  Thus our implementation probably underestimates the speed of their approach.
\item We use the \texttt{u8u16} library~\cite{cameron2008case}. We rely on the latest available version (from 2007). It supports POWER and x64 processors.
\item  We use the \texttt{utf8sse4} library~\cite{Goffart2012}, from the original 2012~release. The entire library  fits in a single source file that has fewer than 300~lines. We corrected a minor bug in the code having to do with the scalar processing, our fix does not impact performance. The library only works on x64 processors supporting the SSE4 instruction set.
\item We use the \texttt{utf8lut} library~\cite{stgatilov}. We use the latest snapshot (last modified April~19, 2020). The library offers a wide range of configuration options that can be passed by C++ template parameters. We benchmark two possible modes: the library provides full  validation any input (\texttt{cmValidate}) and full conversion of any valid input, without validation (\texttt{cmFull}). We leave the other settings to their default values. We experimented with the \texttt{SpeedMult} parameter, changing it from its default value, but we found no obvious performance benefit to non-default settings. The library expects x64 processors.
\end{itemize}
None of the SIMD-accelerated functions we found support ARM processors. As baselines, we also include the following competitors even though they are not focused on SIMD-based acceleration:
\begin{itemize}
\item Unicode Inc.\ makes available Java, C and C++ software providing various Unicode functionality under the name International Components for Unicode (ICU)~\cite{icu}. We use their C++ framework during benchmarking.
Unlike our own work, ICU is comprehensive and provides many more functions than just transcoding ones.
For converting to UTF-16 from UTF-8, we use the 
\texttt{UnicodeString::fromUTF8} function. For converting from UTF-16 to UTF-8, we use their \texttt{UnicodeString::toUTF8String} function.
\item Hoehrmann~\cite{Hoehrmann} published a pure finite-state UTF-8 to UTF-16 transcoder (henceforth \emph{finite}). We use a version of Hoehrmann's transcoder last modified in 2010.
\item Based on Hoehrmann's transcoder, Steagall~\cite{Steagall2018} designed a  fast validating UTF-8 to UTF-16 transcoder. It relies primarily on a finite-state machine with a fast SIMD-based ASCII path. We use the code snapshot they published in October 2018. We made minor changes to improve the portability of the code, but our changes should have no impact on the performance. The library is limited to x64 processors.
\item The LLVM project relies on code routines originally from the Unicode Consortium. They include both UTF-8 to UTF-16 and UTF-16 to UTF-8 transcoding with validation. The code was last revised in September 2001. It does not have any specific processor requirement: it works unchanged under both 64-bit ARM and x64 processors.
\end{itemize}
There are many more C/C++ software libraries providing fast transcoding functions but we expect that we provide a representative set.

It is methodologically difficult to compare software written in different programming languages. Thus we do not include competitive libraries from other programming languages (e.g., Java or Go).

\subsection{Systems}

The performance of our transcoding functions depends on the processor, compiler and other system-specific attributes. To ensure that our results are reasonably robust, we use more than one platform. Our purpose is not, however, to compare the platforms themselves---we do not seek to establish which platform is fastest.

For our  benchmarks, we use a recent x64 processor as well as a recent 64-bit ARM processor from Apple (see Table~\ref{tab:test-cpus}). On the Apple platform, we use Apple's own version of the LLVM clang++ compiler. Our x64 platform runs Linux (Ubuntu) and we rely on the default compiler (GCC). In both cases, we rely on the system ICU library. We use the default flags  provided by the \texttt{cmake} build system in Release---effectively \texttt{-O3}.
A recent AMD or Intel x64 processor has powerful SIMD instructions (256-bit registers or better). ARM processors, such as the Apple~M1, are limited to 128-bit SIMD registers. However, the Apple~M1 processor has more execution units and other favorable attributes that should make it competitive against recent x64 processors.

The x64 processor is located in a server room with appropriate cooling and it is configured for performance. The Apple~M1 processor runs on a laptop. We are mindful of potential methodological issues
when benchmarking on a laptop.
During benchmarking, we record the effective CPU frequency by using performance counters on both platforms: we ensure that the processors reach the maximal frequency during tests for compute-intensive routines. Thus we rule out significant thermal or power constraints. 
\begin{table*}[!tbh]
\caption{\label{tab:test-cpus} Systems tested 
}
\centering\footnotesize
\begin{minipage}{\textwidth}
\centering
\begin{tabular}{cccccc}\toprule
Processor    & Effective Frequency  & Microarchitecture                             & Compiler & ICU version \\ \midrule
 AMD EPYC 7262& \SI{3.39}{\GHz} & Zen~2 (x64, 2019) &  GCC  10.2 & 67.1 \\
Apple M1 & \SI{3.2}{\GHz} & Firestorm (aarch64, 2020) & Apple LLVM  12 & 69.1\\
\bottomrule
\end{tabular}
\end{minipage}
\end{table*}
\subsection{Data}
For a problem such as transcoding, the performance depends on the data.
We collected a wide range of data files in different languages. To ensure reproducibility, we share our data online.\footnote{\url{https://github.com/lemire/unicode_lipsum}} See Table~\ref{table:data}.
\begin{itemize}
\item We use automatically generated (lipsum) text in various languages. The particular files we use were published by Kratz online.\footnote{\url{https://github.com/rusticstuff/simdutf8}} Though the original files were in UTF-8 format, we converted them to UTF-16. The UTF-8 files range from 64KB (Hebrew) to 102KB (Russian).
\item We also recovered the HTML page corresponding to the Mars Wikipedia entry in different languages. The HTML page is in UTF-8. We use the
Python utility \texttt{html2text} to strip the HTML tags and leave only the text of the page.  The UTF-8 file sizes range from 85KB (Esperanto) to 580KB (Thai). 
\end{itemize}
Our data is biased toward natural languages and the basic multilingual plane. Only the Emoji file contains many characters requiring 4-byte UTF-8 sequences. Unsurprisingly, the languages with a higher fraction of 3-byte UTF-8 characters are Chinese, Hindi, Japanese and Korean, Thai. Arabic, Russian and Hebrew have a relatively high fraction of 2-byte UTF-8 characters.
\begin{table}[tb]\centering
\caption{\label{table:data} Quantitative description of our data files. The first two numerical columns represent the average bytes per character in the UTF-8 and UTF-16 formats. The other columns represent the percentage of characters per byte-length in UTF-8 format.}\scriptsize
\subfloat[lipsum]{\restartrowcolors
\begin{tabular}{lcccccc}\toprule
 & UTF-16  &UTF-8 &  1-byte & 2-byte& 3-byte & 4-byte \\\midrule
Arabic & 2.0      &1.8 &  22  & 78  & 0  & 0 \\[10pt]
Chinese  & 2.0      & 3.0& 1  & 0  & 99  & 0 \\[10pt]
Emoji & 4.0      & 4.0 & 0  & 0  & 0  & 100 \\[10pt]
Hebrew & 2.0      & 1.8 & 22  & 78  & 0  & 0 \\[10pt]
Hindi & 2.0      & 2.7 & 16  & 0  & 84  & 0 \\[10pt]
Japanese  & 2.0      & 2.9& 5  & 0  & 95  & 0 \\[10pt]
Korean & 2.0      & 2.5 & 27  & 1  & 72  & 0 \\[10pt]
Latin & 2.0      & 1.0 & 100  & 0  & 0  & 0 \\[10pt]
Russian & 2.0      & 1.8 & 19  & 81  & 0  & 0 \\[10pt]\bottomrule
\end{tabular}
}
\subfloat[Wikipedia Mars]{
   \restartrowcolors
\begin{tabular}{lcccccc}\toprule
 &  UTF-16 &  UTF-8 &1-byte & 2-byte& 3-byte & 4-byte \\\midrule
Arabic & 2.0      & 1.3 & 75  & 25  & 0  & 0 \\
Chinese & 2.0      & 1.3 & 84  & 1  & 16  & 0 \\
Czech & 2.0      & 1.1 & 95  & 5  & 1  & 0 \\
English & 2.0      & 1.0 & 100  & 0  & 0  & 0 \\
Esperanto & 2.0      & 1.0 & 98  & 1  & 1  & 0 \\
French & 2.0      & 1.0 & 98  & 2  & 0  & 0 \\
German & 2.0      & 1.0 & 98  & 1  & 1  & 0 \\
Greek & 2.0      & 1.3 & 74  & 26  & 1  & 0 \\
Hebrew & 2.0      & 1.3 & 71  & 29  & 1  & 0 \\
Hindi & 2.0      & 1.4 & 77  & 0  & 22  & 0 \\
Japanese & 2.0      & 1.4 & 81  & 1  & 19  & 0 \\
Korean & 2.0      & 1.3 & 82  & 1  & 17  & 0 \\
Persan & 2.0      & 1.3 & 76  & 23  & 1  & 0 \\
Portuguese & 2.0      & 1.0 & 98  & 2  & 0  & 0 \\
Russian & 2.0      & 1.3 & 70  & 30  & 0  & 0 \\
Thai  & 2.0      & 1.5& 77  & 0  & 23  & 0 \\
Turkish & 2.0      & 1.1 & 95  & 4  & 1  & 0 \\
Vietnamese & 2.0      & 1.1 & 92  & 4  & 4  & 0 \\\bottomrule
\end{tabular}
}
\end{table}

\subsection{UTF-8 to UTF-16}

In some instances, it may be possible to assume that the UTF-8 input is always valid. Given that the UTF-8 format---unlike the UTF-16 format---requires a non-trivial effort for validation in general, a non-validating transcoder might be faster. We give our results regarding our own implementation as well Inoue et al.\ and \texttt{utf8lut}  in Table~\ref{table:nonvald}. We see that in the absence of a fast path within the Inoue et al.\ approach (i.e., in the Latin case), the performance of our implementation has superior speed. We see that  \texttt{utf8lut} lacks a fast path for ASCII sequences---such an omission could probably be easily remedied. Regarding the Emoji file, the Inoue et al.\ transcoder is not applicable since the file contains characters outside of the basic multilingual plane. Furthermore,  \texttt{utf8lut} has relatively low performance over that file: it falls back on a slower routine when characters outside of the basic multilingual plane are found. Compared to the validating transcoders (see Table~\ref{table:lipsumutf8}), the speed gains of the non-validating approach are often modest for both our approach and \texttt{utf8lut}: e.g., no more than 30\% and sometimes nil. We even see a minute performance regression (5\%) for the Latin dataset due to how the fast ASCII path is implemented.
Though we expect that there are applications where a 30\% speed gain could justify doing away with the validation, we expect that most engineers should rely on a validating transcoder to avoid security concerns.
For the UTF-16 to UTF-8 transcoder, the result is even clearer: there is no measurable benefit to omitting the validation so we omit the numbers.

\begin{table}\centering
\caption{ \label{table:nonvald}Non-validating UTF-8 to UTF-16 transcoding speeds (gigacharacters per second) over the lipsum datasets (AMD Rome (x64), GCC 10)}
\begin{tabular}{lggg}\toprule
   & \mc{Inoue et al.} & \mc{utf8lut} & \mc{ours} \\\midrule
Arabic & 0.29 & 1.2 & 1.5 \\
Chinese & 0.29 & 0.83 & 1.7 \\
Emoji & \mc{unsupported} & 0.24 & 0.53 \\
Hebrew & 0.29 & 1.2 & 1.5 \\
Hindi & 0.29 & 0.94 & 1.2 \\
Japanese & 0.29 & 0.84 & 1.9 \\
Korean & 0.29 & 0.96 & 1.1 \\
Latin & 9.0 & 1.3 & 18. \\
Russian & 0.29 & 1.2 & 1.5 \\\bottomrule
\end{tabular}
\end{table}

We present results regarding the validating UTF-8 to UTF-16 transcoding functions on the lipsum files in  Fig.~\ref{fig:transcoding} as well as in Table~\ref{table:lipsumutf8}.
Focusing on Fig.~\ref{fig:transcoding} where only Arabic, Chinese, Japanese and Korean are presented, 
we find that the LLVM function has the worse performance. ICU, finite, \texttt{u8u16}, \texttt{utf8sse4} have a comparable performance---except on Arabic where the vectorized approaches (\texttt{u8u16} and \texttt{utf8sse4}) are better. Both \texttt{utf8lut} and our approach get close to a gigacharacter per second or better.  Table~\ref{table:lipsumutf8} presents more complete results. 
We find that \texttt{u8u16} and \texttt{utf8sse4} are most competitive on non-Asian languages (e.g., Arabic, Hebrew, Russian). We find
that our approach has superior performance on the Emoji file. Indeed, it appears that other approaches do not optimize the scenario where there are many 4-byte UTF-8 characters.
On the Latin file, we see that only  \texttt{u8u16},  \texttt{utf8sse4} and our approach have a fast path. We find it interesting that finite exceeds a gigacharacter per second on both systems on the Latin file although it lacks a corresponding fast path.
Our approach exceeds a gigacharacter per second on both systems when transcoding Chinese, Hebrew, Japanese, Latin and Hebrew. In these tests, our approach provides an unsurpassed performance: it is typically two times faster than ICU on both systems.

\begin{figure}\centering
\subfloat[AMD system (x64)]{
\includegraphics[width=0.49\textwidth]{lipsumspeed.tikz}
}
\subfloat[Apple M1 system (ARM)]{
\includegraphics[width=0.49\textwidth]{lipsumspeedm1.tikz}
}

\caption{\label{fig:transcoding} Validating UTF-8 to UTF-16 transcoding speed in billions of characters per second for various lipsum files. }
\end{figure}

\begin{table}\centering
\caption{\label{table:lipsumutf8} UTF-8 to UTF-16 validating transcoding speeds (gigacharacters per second) over the lipsum datasets}
\subfloat[AMD Rome (x64), GCC 10]{\scriptsize
\restartrowcolors
\begin{tabular}{lgggggggg}\toprule
    & \mc{ICU} & \mc{LLVM} & \mc{finite} & \mc{Steagall} & \mc{u8u16} & \mc{utf8sse4} & \mc{utf8lut} & \mc{ours} \\\midrule
Arabic & 0.29 & 0.20 & 0.44 & 0.47 & 0.74 & 0.73 & 1.2 & 1.4 \\
Chinese & 0.39 & 0.23 & 0.37 & 0.37 & 0.39 & 0.42 & 0.81 & 1.3 \\
Emoji & 0.15 & 0.18 & 0.25 & 0.23 & 0.27 & 0.16 & 0.21 & 0.47 \\
Hebrew & 0.28 & 0.18 & 0.43 & 0.49 & 0.74 & 0.73 & 1.2 & 1.4 \\
Hindi & 0.34 & 0.19 & 0.27 & 0.28 & 0.43 & 0.48 & 0.92 & 1.0 \\
Japanese & 0.38 & 0.22 & 0.34 & 0.35 & 0.40 & 0.43 & 0.82 & 1.4 \\
Korean & 0.48 & 0.24 & 0.34 & 0.36 & 0.47 & 0.52 & 0.92 & 0.97 \\
Latin & 0.95 & 0.36 & 1.1 & 18. & 11. & 7.2 & 1.3 & 19. \\
Russian & 0.29 & 0.19 & 0.31 & 0.36 & 0.73 & 0.72 & 1.2 & 1.5 \\\bottomrule
\end{tabular}
}
\subfloat[Apple M1 (ARM), LLVM 12]{\scriptsize
\restartrowcolors
\begin{tabular}{lgggg}\toprule
   & \mc{ICU} & \mc{LLVM} & \mc{finite} & \mc{ours} \\\midrule
 & 0.39 & 0.20 & 0.65 & 1.1 \\
 & 0.43 & 0.23 & 0.43 & 1.2 \\
& 0.19 & 0.23 & 0.30 & 0.48 \\
 & 0.39 & 0.20 & 0.68 & 1.1 \\
 & 0.33 & 0.19 & 0.31 & 0.70 \\
 & 0.40 & 0.22 & 0.39 & 1.2 \\
 & 0.39 & 0.19 & 0.49 & 0.76 \\
 & 0.43 & 0.36 & 1.2 & 21. \\
 & 0.32 & 0.20 & 0.38 & 1.1 \\\bottomrule
\end{tabular}
}
\end{table}

Table~\ref{table:marsutf8} presents the results for the wikipedia-Mars datasets. Transcoding these files is relatively easier because they contain more ASCII sequences. We find that \texttt{utf8lut} is often surpassed by \texttt{u8u16} and \texttt{utf8sse4} (English, Esperanto, French, German, Portuguese, Turkish) due to its lack of fast path for ASCII sequences. On the wikipedia-Mars files, \texttt{utf8sse4} is generally better than \texttt{utf8lut}---sometimes by a factor of two. In several cases, it has a performance that matches our own and is even slightly superior in one case (Vietnamese).
Our approach is often four~times faster than ICU with some exceptions (e.g., Korean) where we are merely about twice as fast.

\begin{table}\centering
\caption{ \label{table:marsutf8}UTF-8 to UTF-16 validating transcoding speeds (gigacharacters per second) over the wikipedia-Mars datasets}
\subfloat[AMD Rome (x64), GCC 10]{\scriptsize
\restartrowcolors
\begin{tabular}{lgggggggg}\toprule
   & \mc{ICU} & \mc{LLVM} & \mc{finite} & \mc{Steagall} & \mc{u8u16} & \mc{utf8sse4} &\mc{ utf8lut} &\mc{ours} \\\midrule
Arabic & 0.52 & 0.26 & 0.51 & 0.93 & 1.2 & 1.2 & 1.2 & 2.0 \\
Chinese & 0.65 & 0.28 & 0.59 & 1.2 & 0.98 & 1.4 & 1.1 & 1.4 \\
Czech & 0.73 & 0.29 & 0.68 & 2.0 & 1.3 & 2.0 & 1.1 & 2.2 \\
English & 0.93 & 0.35 & 1.0 & 9.6 & 4.8 & 6.5 & 1.3 & 12. \\
Esperanto & 0.87 & 0.33 & 0.89 & 5.6 & 1.7 & 4.3 & 1.2 & 5.2 \\
French & 0.86 & 0.33 & 0.85 & 3.0 & 1.5 & 2.1 & 1.2 & 2.1 \\
German & 0.89 & 0.34 & 0.91 & 5.7 & 1.8 & 3.8 & 1.2 & 4.0 \\
Greek & 0.53 & 0.26 & 0.53 & 0.94 & 1.2 & 1.4 & 1.2 & 2.4 \\
Hebrew & 0.46 & 0.24 & 0.44 & 0.76 & 1.1 & 1.2 & 1.1 & 1.9 \\
Hindi & 0.54 & 0.26 & 0.48 & 0.83 & 0.94 & 1.1 & 1.1 & 1.3 \\
Japanese & 0.64 & 0.28 & 0.58 & 1.1 & 0.94 & 1.3 & 1.1 & 1.5 \\
Korean & 0.56 & 0.25 & 0.48 & 0.94 & 0.94 & 1.5 & 1.1 & 2.2 \\
Persan & 0.51 & 0.25 & 0.49 & 0.87 & 1.1 & 1.5 & 1.2 & 2.3 \\
Portuguese & 0.87 & 0.33 & 0.86 & 3.5 & 1.5 & 2.0 & 1.2 & 2.4 \\
Russian & 0.49 & 0.24 & 0.46 & 0.83 & 1.1 & 1.1 & 1.2 & 1.6 \\
Thai & 0.64 & 0.30 & 0.61 & 1.1 & 0.94 & 1.1 & 1.1 & 1.6 \\
Turkish & 0.75 & 0.30 & 0.72 & 2.2 & 1.5 & 2.3 & 1.1 & 2.3 \\
Vietnamese & 0.57 & 0.27 & 0.49 & 1.0 & 1.1 & 1.4 & 1.1 & 1.2 \\
\bottomrule
\end{tabular}
}
\subfloat[Apple M1 (ARM), LLVM 12]{\scriptsize
\restartrowcolors
\begin{tabular}{lgggg}\toprule
   & \mc{ICU} & \mc{LLVM} & \mc{finite} & \mc{ours} \\\midrule
 & 0.36 & 0.27 & 0.60 & 2.0 \\
 & 0.38 & 0.28 & 0.66 & 1.6 \\
 & 0.38 & 0.29 & 0.79 & 2.0 \\
& 0.42 & 0.35 & 1.2 & 12. \\
& 0.41 & 0.33 & 1.0 & 4.6 \\
& 0.40 & 0.33 & 0.96 & 2.3 \\
 & 0.41 & 0.34 & 1.1 & 3.7 \\
& 0.36 & 0.27 & 0.62 & 2.1 \\
& 0.34 & 0.25 & 0.52 & 1.7 \\
& 0.36 & 0.26 & 0.56 & 1.5 \\
& 0.38 & 0.28 & 0.65 & 1.6 \\
& 0.35 & 0.24 & 0.56 & 1.6 \\
& 0.36 & 0.26 & 0.60 & 1.9 \\
& 0.41 & 0.33 & 1.0 & 2.8 \\
& 0.35 & 0.26 & 0.56 & 1.7 \\
& 0.39 & 0.29 & 0.70 & 1.8 \\
& 0.39 & 0.30 & 0.84 & 2.5 \\
& 0.34 & 0.27 & 0.57 & 1.3 \\
\bottomrule
\end{tabular}
}
\end{table}

Modern processors allow us to record some performance counters during code execution, with negligible overhead. We record both the  number of instructions retired and the number of instructions per cycle. Table~\ref{table:perfcounter} show that our approach as well as \texttt{utf8lut} distinguish themselves by requiring few instructions per byte. Though the \texttt{utf8lut} approach is sometimes more economical than our approach in terms of instruction count, it leads to code that retires relatively few instructions per cycle. We expect that this limitation is due to the large tables. The ICU approach has both a relatively high instruction count as well as few instructions retired per cycle.
We see that in some cases (e.g., finite) the Apple M1 processor is able to retire more instructions per cycle.
Our approach requires more instructions per byte on the Apple~M1 processor: it is understandable given that the M1 processor has 128-bit SIMD registers compared to 256-bit SIMD registers with the x64 processor.
\begin{table}
    \centering
    \caption{\label{table:perfcounter}Performance-counter measures when transcoding the lipsum Arabic file from UTF-8 to UTF-16}
\subfloat[AMD Rome (x64) processor]{\scriptsize
\restartrowcolors
\begin{tabular}{lgg}
\toprule
 &\mc{ instructions per byte} & \mc{instruction per cycle} \\\midrule
{ICU} &  15  &2.3\\
{LLVM} &  34  &3.5\\
{finite}  &19 & 4.2\\
Steagall  & 17 & 4.2\\
 u8u16  & 10  &3.9\\
 utf8sse4 & 10  &3.9\\
 utf8lut  &3.1 & 2.0 \\
 ours & 4.2 & 3.1 \\ \bottomrule
\end{tabular}
}
\subfloat[Apple M1 (ARM) processor]{\scriptsize
\restartrowcolors
\begin{tabular}{lgg}
\toprule
 &\mc{ instructions per byte} & \mc{instruction per cycle} \\\midrule
{ICU} &  15  &3.3\\[10pt]
{LLVM} &  36  &3.3\\[10pt]
{finite}  &19 & 5.6\\[10pt]
 ours & 6.0 & 3.6 \\[10pt]\bottomrule
\end{tabular}
}
\end{table}
\subsection{UTF-16 to UTF-8}

\begin{figure}
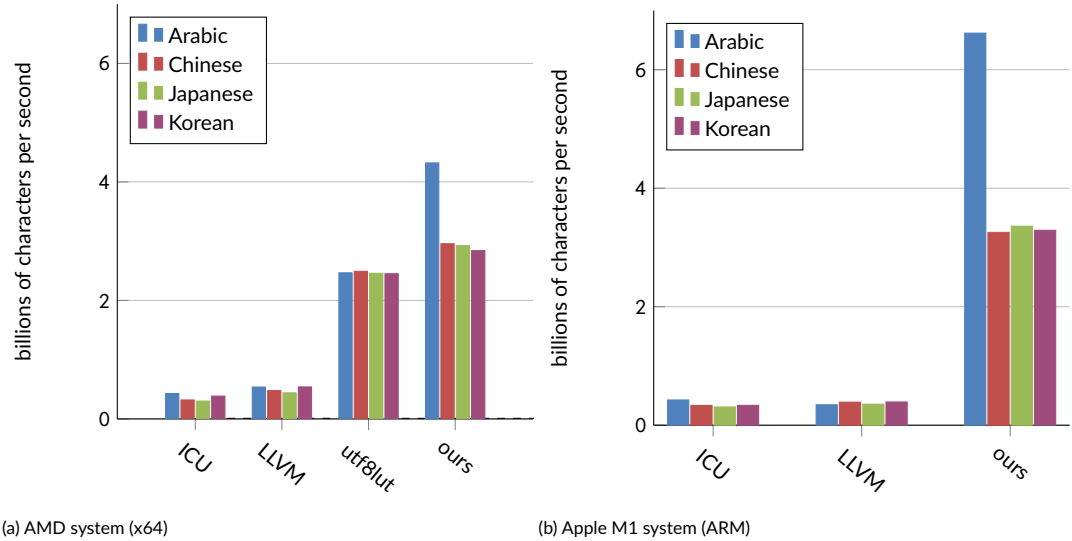
\centering
\subfloat[AMD system (x64)]{
\includegraphics[width=0.49\textwidth]{lipsumspeedutf16.tikz}
}
\subfloat[Apple M1 system (ARM)]{
\includegraphics[width=0.49\textwidth]{lipsumspeedutf16m1.tikz}
}

\caption{\label{fig:transcodingutf16} Validating UTF-16 to UTF-8 transcoding speed in billions of characters per second for various lipsum files. }
\end{figure}

 Fig.~\ref{fig:transcodingutf16}  illustrates the UTF-16 to UTF-8 for some lipsum files. Our approach is between 7~and 10~times faster than ICU\@. Both LLVM and ICU have similar performance, with a slight edge for LLVM---unlike the UTF-8 to UTF-16 results.
  Table~\ref{table:lipsumutf16} provides more complete results over the lipsum files. 
  Our approach has no longer a fast path for transcoding characters outside the basic multiplanar plane. On the AMD Rome system, LLVM and \texttt{utf8lut} offer better performance than our approach on the Emoji file. Otherwise, both ICU and LLVM have far poorer performance than our approach or than \texttt{utf8lut}: the difference is large and often about $10\times$.
  We find it interesting that the performance of our approach on the Apple platform appears to be systematically superior to the performance of our approach on the AMD platform.
  
\begin{table}\centering
\caption{\label{table:lipsumutf16} UTF-16 to UTF-8 validating transcoding speeds (gigacharacters per second) over the lipsum datasets}
\subfloat[AMD Rome (x64), GCC 10]{\scriptsize
\restartrowcolors
\begin{tabular}{lgggg}\toprule
   & \mc{ICU} & \mc{LLVM} & \mc{utf8lut} & \mc{ours} \\\midrule
Arabic & 0.43 & 0.54 & 2.5 & 4.3 \\
Chinese & 0.32 & 0.48 & 2.5 & 3.0 \\
Emoji & 0.23 & 0.40 & 0.40 & 0.33 \\
Hebrew & 0.59 & 0.55 & 2.5 & 4.3 \\
Hindi & 0.24 & 0.36 & 2.5 & 2.9 \\
Japanese & 0.30 & 0.44 & 2.5 & 2.9 \\
Korean & 0.39 & 0.54 & 2.5 & 2.9 \\
Latin & 0.93 & 0.85 & 2.5 & 18. \\
Russian & 0.28 & 0.37 & 2.5 & 4.3 \\
\bottomrule
\end{tabular}
}
\subfloat[Apple M1 (ARM), LLVM 12]{\scriptsize
\restartrowcolors
\begin{tabular}{lggg}\toprule
   & \mc{ICU} & \mc{LLVM} & \mc{ours} \\\midrule
Arabic & 0.43 & 0.35 & 6.6 \\
Chinese & 0.33 & 0.39 & 3.3 \\
Emoji & 0.30 & 0.35 & 0.59 \\
Hebrew & 0.45 & 0.35 & 6.5 \\
Hindi & 0.24 & 0.28 & 3.3 \\
Japanese & 0.31 & 0.35 & 3.3 \\
Korean & 0.34 & 0.39 & 3.2 \\
Latin & 0.77 & 0.40 & 22. \\
Russian & 0.28 & 0.26 & 6.5 \\
\bottomrule
\end{tabular}
}
\end{table}

  The results on the wikipedia-Mars datasets are similar (see Table~\ref{table:marsutf16}). We find that  \texttt{utf8lut} and our approach can be several times faster than LLVM and ICU\@.

\begin{table}\centering
\caption{ \label{table:marsutf16}UTF-16 to UTF-8 validating transcoding speeds (gigacharacters per second) over the wikipedia-Mars datasets}
\subfloat[AMD Rome (x64), GCC 10]{\scriptsize
\restartrowcolors
\begin{tabular}{lgggg}\toprule
   & \mc{ICU} & \mc{LLVM} & \mc{utf8lut} & \mc{ours} \\\midrule
Arabic & 0.43 & 0.52 & 2.5 & 4.8 \\
Chinese & 0.47 & 0.60 & 2.5 & 5.2 \\
Czech & 0.53 & 0.65 & 2.5 & 5.1 \\
English & 0.89 & 0.83 & 2.5 & 14. \\
Esperanto & 0.75 & 0.77 & 2.5 & 10. \\
French & 0.70 & 0.75 & 2.5 & 3.5 \\
German & 0.77 & 0.78 & 2.5 & 8.6 \\
Greek & 0.43 & 0.54 & 2.5 & 6.4 \\
Hebrew & 0.36 & 0.47 & 2.5 & 5.3 \\
Hindi & 0.40 & 0.54 & 2.5 & 5.0 \\
Japanese & 0.47 & 0.60 & 2.5 & 5.2 \\
Korean & 0.36 & 0.54 & 2.5 & 5.3 \\
Persan & 0.40 & 0.52 & 2.3 & 5.2 \\
Portuguese & 0.71 & 0.76 & 2.5 & 4.7 \\
Russian & 0.40 & 0.50 & 2.5 & 4.3 \\
Thai & 0.54 & 0.63 & 2.5 & 4.8 \\
Turkish & 0.57 & 0.68 & 2.5 & 5.3 \\
Vietnamese & 0.39 & 0.53 & 2.5 & 2.8 \\
\bottomrule
\end{tabular}
}
\subfloat[Apple M1 (ARM), LLVM 12]{\scriptsize
\restartrowcolors
\begin{tabular}{lggg}\toprule
    & \mc{ICU} & \mc{LLVM} & \mc{ours} \\\midrule
Arabic & 0.40 & 0.32 & 6.6 \\
Chinese & 0.44 & 0.34 & 6.0 \\
Czech & 0.51 & 0.34 & 9.3 \\
English & 0.74 & 0.40 & 17. \\
Esperanto & 0.63 & 0.38 & 15. \\
French & 0.61 & 0.37 & 6.3 \\
German & 0.66 & 0.38 & 13. \\
Greek & 0.42 & 0.33 & 9.2 \\
Hebrew & 0.35 & 0.30 & 7.9 \\
Hindi & 0.39 & 0.32 & 5.4 \\
Japanese & 0.44 & 0.34 & 5.9 \\
Korean & 0.38 & 0.31 & 6.8 \\
Persan & 0.40 & 0.32 & 9.6 \\
Portuguese & 0.62 & 0.37 & 9.3 \\
Russian & 0.38 & 0.31 & 6.4 \\
Thai & 0.47 & 0.36 & 5.5 \\
Turkish & 0.54 & 0.34 & 8.7 \\
Vietnamese & 0.40 & 0.31 & 3.7 \\
\bottomrule
\end{tabular}
}
\end{table}

\subsection{Varying the Input Size}
\begin{figure}
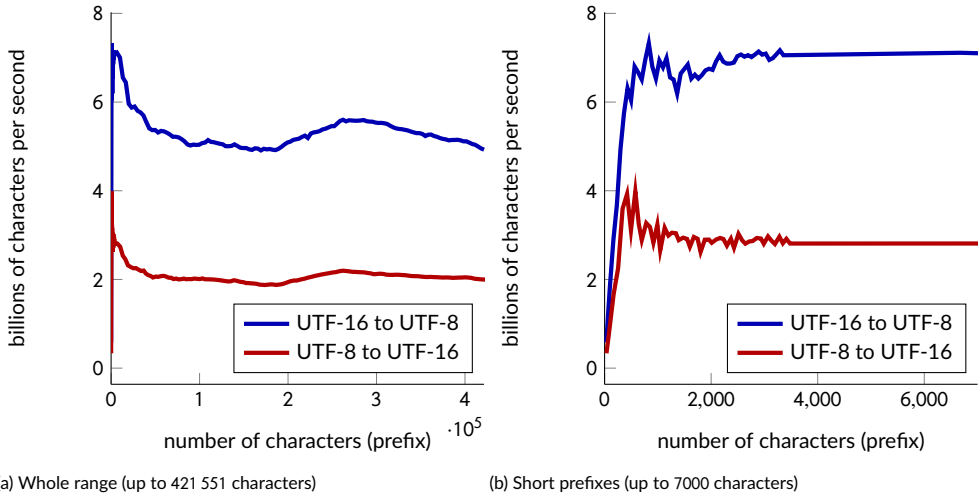
\centering
\subfloat[Whole range (up to \num{421551}~characters)]{
\includegraphics[width=0.45\textwidth]{wikipediaarabic.tikz}
}
\subfloat[Short prefixes (up to \num{7000}~characters)]{
\includegraphics[width=0.45\textwidth]{wikipediaarabicshort.tikz}
}
\caption{\label{fig:transcodingutf16variouslengths} Validating transcoding speed in billions of characters per second for prefixes of various lengths---from 1~to  to \num{421551}~characters---of the Arabic Wikipedia Mars data file (AMD Rome system) using our techniques. }
\end{figure}

One might be concerned that our good results depend on the size of the string to be transcoded. And, indeed, over tiny strings, there is little benefit to be expected from SIMD instructions. However, Fig.~\ref{fig:transcodingutf16variouslengths} shows that as soon as we go beyond $\approx 100$~bytes, the performance of our transcoders quickly reaches speeds in the gigacharacters per second range. We constructed the plots by benchmarking both the  UTF-8 to UTF-16  and UTF-16 to UTF-8 transcoding speeds on longer and longer prefixes of the Arabic lipsum file. After constructing the prefix---either in UTF-8 or UTF-16 format---we transcode it as many times as needed for a total of at least \SI{0.2}{\second}. For these multiple runs, we deduce the speed. 

On small prefixes, because we repeatedly process the same small string, we may underestimate the cost of branch mispredictions. Nevertheless, our results remain an indication that we can get high speed on relatively small strings.

\subsection{Discussion}

The problems of transcoding from UTF-8 and UTF-16, and transcoding from UTF-16 to UTF-8 appears roughly symmetrical in performance with conventional functions (e.g., ICU and LLVM).
Our SIMD-accelerated functions break this symmetry.
When comparing the number of characters transcoded by second, we find that 
transcoding UTF-16 to UTF-8 is faster  than transcoding  UTF-8 to UTF-16---sometimes by a factor of two or more. Intuitively, we might expect such a result since decoding UTF-8 is a more demanding task due to its higher variability in character length (1, 2, 3~or 4~bytes per character). Most UTF-16 strings are mere sequences of 2-byte characters. Validating UTF-16 is also simpler. Yet it is possible that our UTF-8 to UTF-16 functions are not as efficient as they could be.

Gatilov's \texttt{utf8lut} library is competitive with our own work. However, it does a different trade-off with respect to memory usage. It requires a sizeable table \SI{2}{\mebi\byte} for UTF-8 to UTF-16 as opposed to about \SI{11}{\kibi\byte} in our case. 
As an intermediate case,
Inoue et al.~\cite{Inoue2008} require a table of about \SI{105}{\kibi\byte} for their UTF-8 to UTF-16 transcoder.

For UTF-16  to UTF-8, Gatilov's \texttt{utf8lut} library requires a table of 256~64-byte entries (\SI{16}{\kibi\byte}) for the basic multilingual plane. We need two tables of 256~17-byte entries (\SI{8.5}{\kibi\byte}) for the same task. 

Large tables may not necessarily be a practical concern, but it is important to know whether they are necessary for high performance. Our results suggest that large tables are unnecessary. However, future work could explore further the tradeoffs.

Though it only supports UTF-8 to UTF-16 transcoding under recent x64 processors, the \texttt{utf8sse4} approach can sometimes surpass  Gatilov's \texttt{utf8lut} library. When there are many ASCII sequences, it can even match our own performance.
The \texttt{u8u16} library is similarly competitive. However,  \texttt{utf8sse4} and  \texttt{u8u16} may generate more than twice the number of instructions compared with our own approach or \texttt{utf8lut} (see Table~\ref{table:perfcounter}).

\section{Conclusion}

Our work demonstrates that validating SIMD-based transcoders can surpass conventional transcoders (e.g., ICU) by a wide margin (e.g., $4\times$) over many data sources. As we demonstrate by our implementations and algorithms, these benefits do not require large inputs, particular data alignment, specialized hardware or code bloat (large tables).


Future work could investigate interleaved decoding. Given a long string, one could decode the first half and the second half separately---for example. One needs to ensure that the outputs end up being consecutive which we can achieve by copying them or by pre-computing the character offsets.

Our approach accelerate the full Unicode set when transcoding from UTF-8 to UTF-16. Thus we provide a high speed even in the presence of emojis---unlike most other competing techniques. However, our UTF-16 to UTF-8 transcoding functions only accelerates the basic multilingual plane. Future work should seek to lift this limitation without introducing large tables.

\bibliography{utf16utf8}

\end{document}